\documentclass[prd,twocolumn,superscriptaddress]{revtex4}
\usepackage{amsmath,amssymb,epsfig,bm,mathrsfs,feynmp,color}
\usepackage{slashed}
\newcommand{\be}{\begin{equation}}
\newcommand{\ee}{\end{equation}}
\newcommand{\bea}{\begin{eqnarray}}
\newcommand{\eea}{\end{eqnarray}}

\newcommand{\de}{\partial}

\newcommand{\vecq}{{\mathbf q}}
\newcommand{\vecp}{{\mathbf p}}

\newcommand{\vectau}{{\bm \tau}}
\newcommand{\vecgamma}{{\bm \gamma}}

\newcommand{\vecpi}{{\bm \pi}}
\newcommand{\ie}{{\it i.e.}}

\def\Xint#1{\mathchoice
   {\XXint\displaystyle\textstyle{#1}}%
   {\XXint\textstyle\scriptstyle{#1}}%
   {\XXint\scriptstyle\scriptscriptstyle{#1}}%
   {\XXint\scriptscriptstyle\scriptscriptstyle{#1}}%
   \!\int}
\def\XXint#1#2#3{{\setbox0=\hbox{$#1{#2#3}{\int}$}
     \vcenter{\hbox{$#2#3$}}\kern-.5\wd0}}

\def\dashint{\Xint-}

\definecolor{red}{rgb}{0.8,0,0}
\definecolor{orange}{rgb}{0.8,0.2,0.0}
\definecolor{blue}{rgb}{0.3,0.0,0.8}

\begin{document}

%%%%%%%%%%%%%%%%%%%%%%%%%%%%%%%%%%%%%%%%%%%%%%%%%%%%

\title{Chiral pions in a magnetic background}

\author{Giuseppe Colucci} 
\affiliation{Institute for Theoretical
   Physics, J.~W.~Goethe-University, D-60438 Frankfurt-Main, Germany}

\author{Eduardo S. Fraga}
\affiliation{Institute for Theoretical
   Physics, J.~W.~Goethe-University, D-60438 Frankfurt-Main, Germany}
\affiliation{Instituto de F\'isica, Universidade Federal do Rio de Janeiro,
  Caixa Postal 68528, Rio de Janeiro, RJ 21945-970, Brazil}

\author{Armen Sedrakian} 
\affiliation{Institute for Theoretical
   Physics, J.~W.~Goethe-University, D-60438 Frankfurt-Main, Germany}

%%%%%%%%%%%%%%%%%%%%%%%%%%%%%%%%%%%%%%%%%%%%%%%%%%%%
\begin{abstract}
We investigate the modification of the pion self-energy at finite temperature 
due to its interaction with a low-density, isospin-symmetric nuclear medium embedded in a 
constant magnetic background. To one loop, for fixed temperature and density, we find that 
the pion effective mass increases with the magnetic field. For the $\pi^{-}$, interestingly, this happens 
solely due to the trivial Landau quantization shift $\sim |eB|$, since the real part of the 
self-energy is negative in this case. In a scenario in which other charged particle species are present 
and undergo an analogous trivial shift, the relevant behavior of the effective mass might be determined 
essentially by the real part of the self-energy. In this case, we find that the pion mass decreases by
$\sim 10\%$ for a magnetic field $|eB|\sim m_\pi^2$, which favors pion condensation at high density and low 
temperatures.  
\end{abstract}

% \pacs{03.75.Fi, 05.30.Jp, 67.40.Db, 67.40.Vs}

\maketitle

%%%%%%%%%%%%%%%%%%%%%%%%%%%%%%%%%%%%%%%%%%%%%%%%%%%%
\section{Introduction}

The behavior of hadronic matter in a medium under the influence of a strong external magnetic field 
can be very rich and subtle, and has been the subject of intense investigation in the last few years. 
In fact, in-medium strong interactions under extreme magnetic fields are of experimental relevance in 
heavy ion collisions and in astrophysics, exhibit a rich new phenomenology and are amenable to lattice 
simulations. (For comprehensive reviews, see Ref. \cite{Kharzeev:2013jha}.)

Even if every model calculation has predicted that large enough magnetic fields, typically of the order of a few times 
$m_{\pi}^2$, could bring remarkable new features to the thermodynamics of strong interactions, from shifting the chiral and the 
deconfinement crossover lines in the phase 
diagram \cite{Agasian:2008tb,Fraga:2008qn,Menezes:2008qt,Boomsma:2009yk,Mizher:2010zb,Fukushima:2010fe,Gatto:2010pt,Kashiwa:2011js,Chatterjee:2011ry,Andersen:2011ip,Skokov:2011ib,Andersen:2012bq,Fukushima:2012xw,Fraga:2012fs} 
to transforming the vacuum into a superconducting medium via $\rho$-meson condensation \cite{Chernodub:2010qx,Chernodub:2011mc}, 
essentially all models fail to describe coherently the available lattice data \cite{D'Elia:2010nq,D'Elia:2011zu,Bali:2011qj,Bali:2012zg}. The reasons for 
that are still unclear, although there are some indications that confinement plays a relevant role \cite{Fraga:2012fs,Fraga:2012ev}, which 
is not captured in the usual low-energy effective chiral models of QCD~\cite{Fraga:2012rr}. 
In any case, the situation calls for theoretical investigations in more 
controlled setups, with less freedom and parameters to adjust. This approach has proved to be fruitful in the 
large-$N_{c}$ \cite{Fraga:2012ev} and perturbative \cite{Blaizot:2012sd} limits of QCD: in the former, the behavior of the critical 
temperature for deconfinement was found to be in qualitative agreement with lattice data; in the latter, a trivial chiral limit 
for the two-loop contribution to the QCD pressure in a strong magnetic background was revealed.

Following this line of action, a natural extension is the study of hadronic matter in the complementary, low-energy sector, 
in the presence of a strong magnetic field, in a controlled setup. Thus, since we are interested in the low-density, low-temperature 
sector of the phase diagram of nuclear matter, we adopt the framework of chiral perturbation theory, which represents 
a powerful tool to study the low-energy regime of the pion-nucleon physics~\cite{2011PhR...503....1M}. 

It is the purpose of this work to investigate 
some properties of isospin-symmetric nuclear matter in the limit of low density and temperature, embedded in a strong 
magnetic background. In particular, we study the modifications of the spectrum of the lowest energy degree of freedom, 
the pion, due to the interaction with nucleons and the constant magnetic field. More specifically, we compute the pion effective 
mass in the presence of a constant magnetic field to one loop. (Even if we do not address the phase diagram here, it should 
be mentioned that the inclusion of nucleons, and pion-nucleon interactions, proved to be necessary for a satisfactory description 
of the behavior of the deconfinement critical temperature as a function of the pion mass and isospin \cite{Fraga:2008be}.) 
For this purpose, we consider fully relativistic chiral perturbation theory as a framework for our computation. This is
needed to define consistently the fermion propagators in a magnetic background. At the same time, 
this work extends a previous treatment on the calculation of the fermion self-energy in relativistic 
chiral perturbation theory~\cite{2013PhRvC..88a5209C}.

In-medium pion properties have been extensively investigated, both in 
finite systems, {\ie} pionic atoms \cite{1997PhLB..405..215W, 2003PhRvL..90i2501K}, and in
infinite nuclear matter. In the latter, an interesting aspect of pion phenomenology 
is represented by pion condensation at high densities, introduced by Migdal 
\cite{1973PhRvL..31..257M}, which is a consequence of the fact that at high density the
electron chemical potential grows until it is favorable for a neutron on the 
top of the Fermi sea to turn into a proton and a (negatively charged) pion. 
On other hand, the interaction of the pion with the background matter 
can enhance its self-energy and consequently the pion condensation threshold density. 
This issue is still open and requires more investigation because of its 
implications in the context of compact stars 
phenomenology \cite{2009PhRvC..80c8202O,2011ApJS..197...20S,2013PhRvD..87d3006P}. 
We shall see in the sequel that the in-medium modification of the (negatively charged) pion self-energy due to the 
presence of a strong magnetic background might lead to relevant phenomenological consequences.

The paper is organized as follows. In Section~\ref{section:reminder}, we consider the relativistic formulation 
of the theory, since in this framework it is possible to define the Green's function of the theory in the presence of a 
constant magnetic background in a consistent fashion. In Section~\ref{section:self-energy},
we compute the lowest order pion self-energy for the three charge eigenstates in isospin 
symmetric nuclear matter. In Section~\ref{section:results}, we compute the in-medium effective 
mass of the pion and its dependence on the value of the applied magnetic field. Finally, in 
Section~\ref{section:conclusions}, we summarize our conclusions.
We use natural units $\hbar = c = k_B = 1$. Four-vectors are denoted by capital letters, 
for instance $P^\mu = (p_0,\vecp)$.

%%%%%%%%%%%%%%%%%%%%%%%%%%%%%%%%%%%%%%%%%%%%%%%%%%%%
\section{Reminder of the pion effective mass}
\label{section:reminder}

The low-energy phenomenology of pions in nuclear matter is well described in terms 
of a chirally invariant pion-nucleon interaction Lagrangian, expanded in powers 
of the low-energy scale of the theory, {\ie} the ratio of the pion momentum or 
mass over ($4\pi$ times) the pion decay constant:
\be\label{eq:chiral_lagrangian}
  {\cal L}_{\pi N} = {\cal L}^{(1)}_{\pi N} + {\cal L}^{(2)}_{\pi N}+...
\ee
where the leading order, ${\cal L}_{\pi N}^{(1)}$, reads \cite{2011PhR...503....1M}
\be
  {\cal L}_{\pi N}^{(1)} = - \bar\Psi\left[%i\gamma^\mu\de_\mu - m 
                           \frac{g_A}{2f_\pi}\gamma^\mu\gamma_5\vectau\cdot\de_\mu\vecpi
                           +\frac{1}{4 f_\pi^2}\gamma^\mu\vectau\cdot(\vecpi\times\partial_\mu\vecpi)
                           \right]\Psi.\label{eq:LO_lagrangian} 
\ee
Here $\vectau$ is the vector of Pauli matrices in isospin space, $\vecpi$ is
the isotriplet of pions, $f_\pi$ the pion decay constant and $g_A$ is the axial-vector 
coupling.
% , whose value is determined from neutron beta decay and is given by 
% $g_A = 1.2695 \pm 0.0029$. 
\begin{figure}[th]
 \centering
 \includegraphics[keepaspectratio=true]{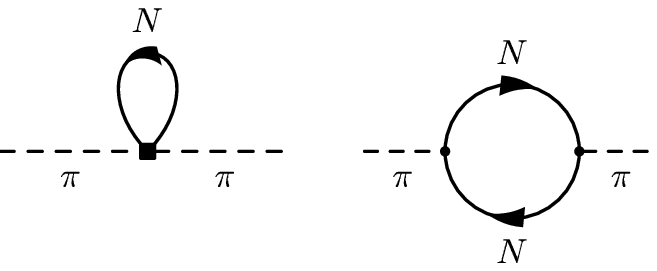}
 % feynman.eps: 0x0 pixel, 300dpi, 0.00x0.00 cm, bb=148 615 349 677
 \caption{Diagrams contributing to the lowest-order in-medium pion self-energy. 
 The small dotted vertex corresponds to the one-pion exchange part of the Lagrangian 
 in Eq.~(\ref{eq:LO_lagrangian}), while the squared one to the two-pion exchange in the 
 Weinberg-Tomozawa term.}
 \label{fig:diagrams}
\end{figure}

The diagrams contributing to the pion self-energy from the Lagrangian~(\ref{eq:LO_lagrangian})
are shown in Fig.~\ref{fig:diagrams}. The former is obtained from the Weinberg-Tomozawa 
term, while the latter comes from the one-pion exchange Lagrangian. 
Due to the coupling of the charge to the vector potential,
in the case of a constant magnetic background we need to compute separately those diagrams 
for different pion and nucleon charge eigenstates.

\begin{figure}[ht]
 \centering
 \includegraphics[keepaspectratio=true]{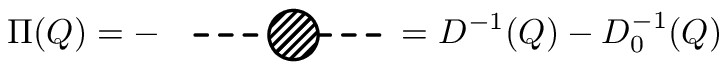}
 % feynman.eps: 0x0 pixel, 300dpi, 0.00x0.00 cm, bb=148 615 349 677
 \caption{Pion Schwinger-Dyson equation. Here $D_0$ is the free pion propagator and $D$ is the full one. 
 The diagram in the previous equation denotes the sum of all one-particle irreducible (1PI) diagrams. 
 $Q^\mu = (\omega,\vecq)$ is the pion four momentum.}
 \label{fig:self-energy-pics}
\end{figure}

Formally, the self-energy can be defined from the pion Schwinger-Dyson equation \cite{LeBellac}, 
displayed in Fig.~\ref{fig:self-energy-pics}. 
% In this figure $D_0$ is the free pion propagator and $D$ is the full one. The diagram in the previous 
% equation denotes the sum of all one-particle irreducible (1PI) diagrams. $Q^\mu = (\omega,\vecq)$
% is the pion four momentum.
% At finite temperature and density, $Q^\mu$ is a four-vector and its zeroth component is 
% defined as $q^0=iq_n + \mu$, $iq_n= 2n\pi i T $, $n\in \mathbb{Z}$ and $T$ is the temperature.

Pionic modes of excitation in nuclear matter are obtained as solutions $\omega(\vecq)$ 
of the following equation
\be\label{eq:dispersion_relation}
  \omega^2 - \vecq^2 - m_\pi^2 + \Pi(\omega,\vecq) = 0\,,
\ee
and in the limit of vanishing momenta this solution corresponds to the effective pion mass
\be\label{eq:eff_mass}
  {m_\pi^*}^2 = m_\pi^2 - {\rm Re}~\Pi({m_\pi^*},\vecq=0)\,.
\ee

In absence of a magnetic background, it can be shown that the lowest-order (LO) 
contribution to the effective mass in (\ref{eq:eff_mass}) vanishes in 
isospin symmetric nuclear matter \cite{2011AnPhy.326..241L}. 
% ,pion_mass_preparation

In asymmetric nuclear matter, the LO self-energy of the (negatively charged) pion receives a contribution
from the Weinberg-Tomozawa diagram, given by \cite{PhysRevLett.90.092501} 
\be
  \Pi^{\rm WT}(\omega,\vecq=0) = \frac{\omega}{2 f_\pi^2}(\rho_p - \rho_n)\,.
\ee

In the presence of a magnetic background, the pion charge eigenstates 
Eq.~(\ref{eq:eff_mass}) have to be modified (due to the Landau level quantization) to 
\be\label{eq:eff_mass_magnetic}
  {m_\pi^*}^2 = m_\pi^2 - {\rm Re}~\Pi({m_\pi^*}^2,\vecq=0;\mathbf B)\, + (2n + 1)|eB|\,,
\ee
where $\mathbf B$ is the magnetic field and $n$ is the index of the Landau level. 

In what follows we focus on the case of symmetric nuclear matter in the presence 
of a constant magnetic background. Thus, any deviation from zero of the LO pion self-energy 
will give a contribution to the effective pion mass in a magnetic background. Since we 
are dealing with dilute nuclear matter at low temperatures, we neglect the contribution
of anti-nucleons. Moreover, we choose the $x_3-$axis to be parallel to the magnetic field and $|eB| = eB$, 
$e$ being the proton electric charge. In order to simplify the calculation, we assume the regime of 
strong magnetic fields, in which one can apply the lowest-Landau-level (LLL) approximation to 
simplify the propagators. We neglect the effect of the anomalous 
magnetic moment of protons and neutrons. The calculation is carried out in 
the Landau gauge.

%%%%%%%%%%%%%%%%%%%%%%%%%%%%%%%%%%%%%%%%%%%%%%%%%%%%
\section{Pion self-energy in a constant magnetic field}
\label{section:self-energy}

For the negatively charged pion, the first diagram in Fig.~\ref{fig:diagrams} 
leads to the following contribution:
\be\label{eq:WT_self}
  \Pi^{\rm WT}(Q) = \Pi^{\rm WT}_p(Q) - \Pi^{\rm WT}_n(Q)\,,
\ee
where the first term is the proton loop contribution, which by using the Furry representation 
at finite temperature for the proton propagator \cite{1996NuPhB.479....3E} reads
\be\label{eq:WT_self_p}
  \Pi^{\rm WT}_p(Q) = \frac{1}{f_\pi^2}\frac{|eB|}{4\pi^2}
                          \int_{-\infty}^{+\infty}dp_3~
                          n_F(E_3-\mu)\frac{\vecp_L\cdot\vecq_L}{2E_3}\,,
\ee
whereas the neutron loop contribution is not affected by the presence of the magnetic field
\be\label{eq:WT_self_n}
  \Pi^{\rm WT}_n(Q) = \frac{1}{f_\pi^2}\int\frac{d^3\vecp}{(2\pi)^3}
                         \frac{\omega E_p - \vecp\cdot\vecq}{E_p} n_F(E_p - \mu) \,.
\ee
In Eq.~(\ref{eq:WT_self_p}) $n_F(x) = (e^{x/T}+1)^{-1}$ is the Fermi distribution, 
$E_3 = \sqrt{p_3^2 + m^2}$, $m$ being the proton mass, and the subscript $L$ indicates that 
the vectors live in the two-dimensional subspace defined by the time component and the
space component that is aligned with the magnetic field, {\ie} $\vecp_L = (p_0,p_3)$.

The WT self-energy for the positively charged pion will be just the opposite of Eq.~(\ref{eq:WT_self}). 
Finally, the $\pi^0$ does not receive any one-loop contribution from the WT interaction term.

The contribution to the pion self-energy from the one-pion exchange term in 
the Lagrangian (\ref{eq:LO_lagrangian}) is, on the other hand, quite involved. For the charged
pion one has that the two nucleons in the diagram correspond to two different isospin 
states, one being a proton and the other a neutron. For the proton we choose now a more convenient
form for the propagator \cite{1951PhRv...82..664S}, which in the LLL approximation reads
\be
\begin{split}
  S^{(p)}_{LLL}(X,Y) & = \int~\frac{dp_0dp_2dp_3}{(2\pi)^3}\\
                     & \sqrt{\frac{|eB|}{\pi}}e^{-ip_0(x_0-y_0)+ip_2(x_2-y_2)+ip_3(x_3-y_3)}\\
                     & \exp\left\{-\frac{|eB|}{2}\left[\left(x_1-\frac{p_2}{eB}\right)^2+\left(y_1-\frac{p_2}{eB}\right)^2\right]\right\}\\
                     & {\cal P}_0\frac{\vecp_L\cdot\vecgamma_L + m}{\vecp_L^2-m^2}\,,
                     \label{eq:schwinger-propagator}
\end{split}
\ee
where ${\cal P}_0 = \frac{1}{2}[1-i\gamma^1\gamma^2{\rm sign}(eB)]$.
The self-energy in the coordinate space is given by
\be\label{eq:OPE-self-energy}
  \Pi^{\rm OPE}(X,Y) = -\frac{g_A^2}{f_\pi^2}{\rm Tr}\left[\gamma_5\gamma^\mu S_p(X,Y)\gamma_5\gamma^\nu S_n(Y,X)Q_\mu Q^\prime_\nu\right]\,,
\ee
where $Q_\mu$ and $Q^\prime_\nu$ are defined in the momentum space as the pion momenta
\be\label{eq:fourier}
  \Pi^{\rm OPE}(Q,Q^\prime) = \int~d^4X~d^4Y e^{i(Q^\prime\cdot X + Q\cdot Y)}\Pi^{\rm OPE}(X,Y)\,.
\ee
By substituting the propagator in Eq.~(\ref{eq:schwinger-propagator}), one can write
\be
  \Pi^{\rm OPE}(Q,Q^\prime) = (2\pi^3)\delta^{(0,2,3)}(q^\prime +q) \tilde\Pi^{\rm OPE}(Q), 
\ee
where
\be\label{eq:ope-result-T0}
\begin{split}
  \tilde\Pi^{\rm OPE}(Q) & =  \frac{g_A^2}{f_\pi^2}\sqrt{\frac{16\pi}{|eB|}}Q_\mu Q_\nu\int\frac{d^4P}{(2\pi)^4}\\
                         & \times  \frac{F^{\mu\nu}(P)}{(P^2-m^2)[(\vecp +\vecq)_L^2-m^2]}\\
                         & \times  \exp\left\{-\frac{1}{eB}\left[(q_1+p_1)^2\right]\right\}
\end{split}
\ee
and 
\be
  F^{\mu\nu}(P) = (\vecp +\vecq)_L^\mu P^\nu + P^\mu(\vecp +\vecq)_L^\nu - g^{\mu\nu}[(\vecp +\vecq)_L\cdot\vecp_L + m^2]\,.
\ee
Notice that, in Eq.~(\ref{eq:ope-result-T0}), we used the fact that the $\delta$-function and the on-shell condition 
for the pion lead to $q_1+q_1^\prime = 0$.

Since the pole structure of the propagators was not modified, the self-energy at 
finite temperature is obtained by replacing the time component of the four momenta by the 
appropriate (fermionic or bosonic) Matsubara frequency, {\ie} $P^\mu = (p_0,\vecp)$, with $p^0=ip_n + \mu$, 
$ip_n= 2n\pi i T $ for the pion and $ip_n= (2n+1)\pi i T $ for the nucleon, $n\in \mathbb{Z}$ and $T$ being the temperature.

Performing the sum over the fermionic Matsubara frequency, the retarded self-energy reads
\be
\begin{split}
  \tilde\Pi^{\rm OPE}(Q) & = - \frac{g_A^2}{f_\pi^2}\sqrt{\frac{16\pi}{|eB|}}Q_\mu Q_\nu \int_{-\infty}^\infty
  \frac{d^3\vecp}{(2\pi)^3}e^{-\frac{1}{eB}(q_1+p_1)^2}\\
  &\times \frac{1}{4 E_p E^L_{pq}} \bigg[F^{\mu\nu}(E_p,\vecp) n_F(E_p - \mu) \\
  & ~~~~~~~- F^{\mu\nu}(E^L_{pq} - q_0,\vecp) n_F(E^L_{pq} - \mu)\bigg]\\
  &\times \frac{1}{q_0 + E_p - E^L_{pq} + i\eta}\,,
  \label{eq:full_ope}
\end{split}
\ee
where $E^L_{pq} = \sqrt{(\vecp_L - \vecq_L)^2 + m^2}$.
Notice that in this case, due to the presence of a neutral field in the loop, one does not 
have the dimensional reduction which takes place in the case of the gluon self-energy 
in QCD \cite{1996NuPhB.462..249G,2011PhRvD..83k1501F,2012PhDT........17P}.
This is instead the case for the $\pi^0$, as we shall see in the following.

We separate the imaginary and real parts of the self-energy in 
Eq.~(\ref{eq:full_ope}) by means of the Sokhotski-Plemelj formula.
% \be
%   \frac{1}{\omega - \omega_0 + i\eta} = 
%   P.V.\left\{\frac{1}{\omega - \omega_0}\right\} - i\pi\delta(\omega - \omega_0)\,.
% \ee
Since the imaginary part of the self-energy satisfies the hypothesis 
of the Kramers-Kronig dispersion relation, the real part can be computed as
\be
  {\rm Re}\,\Pi^{\rm OPE}(\omega,\vecq) = 
         \frac{1}{\pi}
%          {\rm P.V.}\int
         \dashint_{-\infty}^{\infty}d\omega^\prime
         \frac{{\rm Im}\,\Pi^{\rm OPE} (\omega^\prime,\vecq)}{\omega^\prime-\omega}. 
\ee
As before, the result for the positively charged pion will be the opposite as compared to the $\pi^-$.

The one-loop contribution to the neutral pion self-energy is the sum of the proton and the neutron loops.
The neutron loop is not affected by the presence of the magnetic field and vanishes in the limit of $\vecq\rightarrow0$.
The proton loop can be computed in a way that is very similar to the charged pion self-energy computation. The result reads
\be
\begin{split}
  \Pi_0(Q,Q^\prime) & = - \frac{g_A^2}{2 f_\pi^2}(2\pi)^4\delta^4(Q^\prime + Q)\frac{|eB|}{2\pi}
  e^{-\frac{q_1^2 + q_2^2}{2|eB|}}Q_\mu Q_\nu\\
  &\times \int\frac{dp_3}{2\pi}\frac{1}{4 E^L_p E^L_{pq}}\bigg[H^{\mu\nu}(E^L_p,p_3)n_F(E^L_p - \mu)\\ 
  & ~~~~~~~~~~~- H^{\mu\nu}(E^L_{pq} - iq_n,p_3)n_F(E^L_{pq} - \mu)\bigg]\\
  &\times\frac{1}{E^L_p - E^L_{pq} + iq_n}\,,\label{eq:pi_0_self}
\end{split}
\ee
where 
\be
\begin{split}
  H^{\mu\nu}(\vecp_L) & = \vecp_L^\mu(\vecp_L+ \vecq_L)^\nu + \vecp_L^\nu(\vecp_L+ \vecq_L)^\mu\\
                      & - g^{\mu\nu}[\vecp_L\cdot(\vecp_L+ \vecq_L) + m^2]\,.
\end{split}
\ee
As already stated previously, in this case we recover the dimensional reduction, from (3+1) to (1+1),
where the only spatial dimension that is relevant for the dynamics is determined by the direction of the magnetic field.

%%%%%%%%%%%%%%%%%%%%%%%%%%%%%%%%%%%%%%%%%%%%%%%%%%%%
\section{Results}
\label{section:results}

We compute numerically the integrals appearing in Eqs.~(\ref{eq:WT_self}),~(\ref{eq:full_ope}) and~(\ref{eq:pi_0_self}).
In the limit of $\vecq\rightarrow0$ the only nonvanishing contribution to the self-energy is given by the
WT contribution in Eq.~(\ref{eq:WT_self}). This is due to the fact that in symmetric nuclear matter
the poles in Eq.~(\ref{eq:full_ope}) and (\ref{eq:pi_0_self}) vanish for $\vecq\rightarrow0$, leading thus to a
vanishing 
% imaginary and, consequently, real part of the 
self-energy. So, the neutral pion, whose self-energy
comes only from the one-pion exchange term in the Lagrangian, has its mass unaltered as is also the case  
when no magnetic background is present.

\begin{figure}[htb]
 \centering
 \includegraphics[keepaspectratio=true,width=0.5\textwidth]{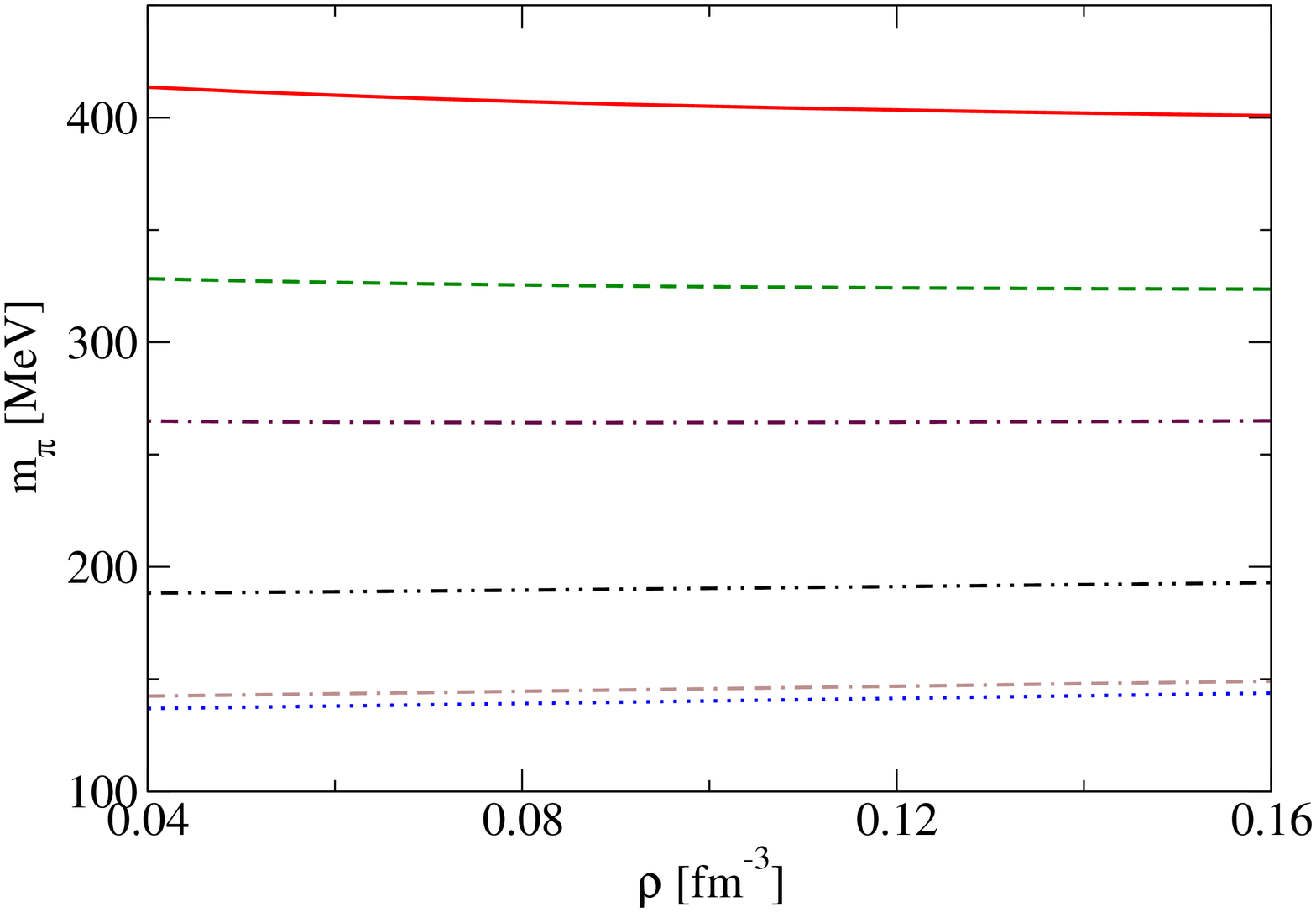}
 
 \includegraphics[keepaspectratio=true,width=0.5\textwidth]{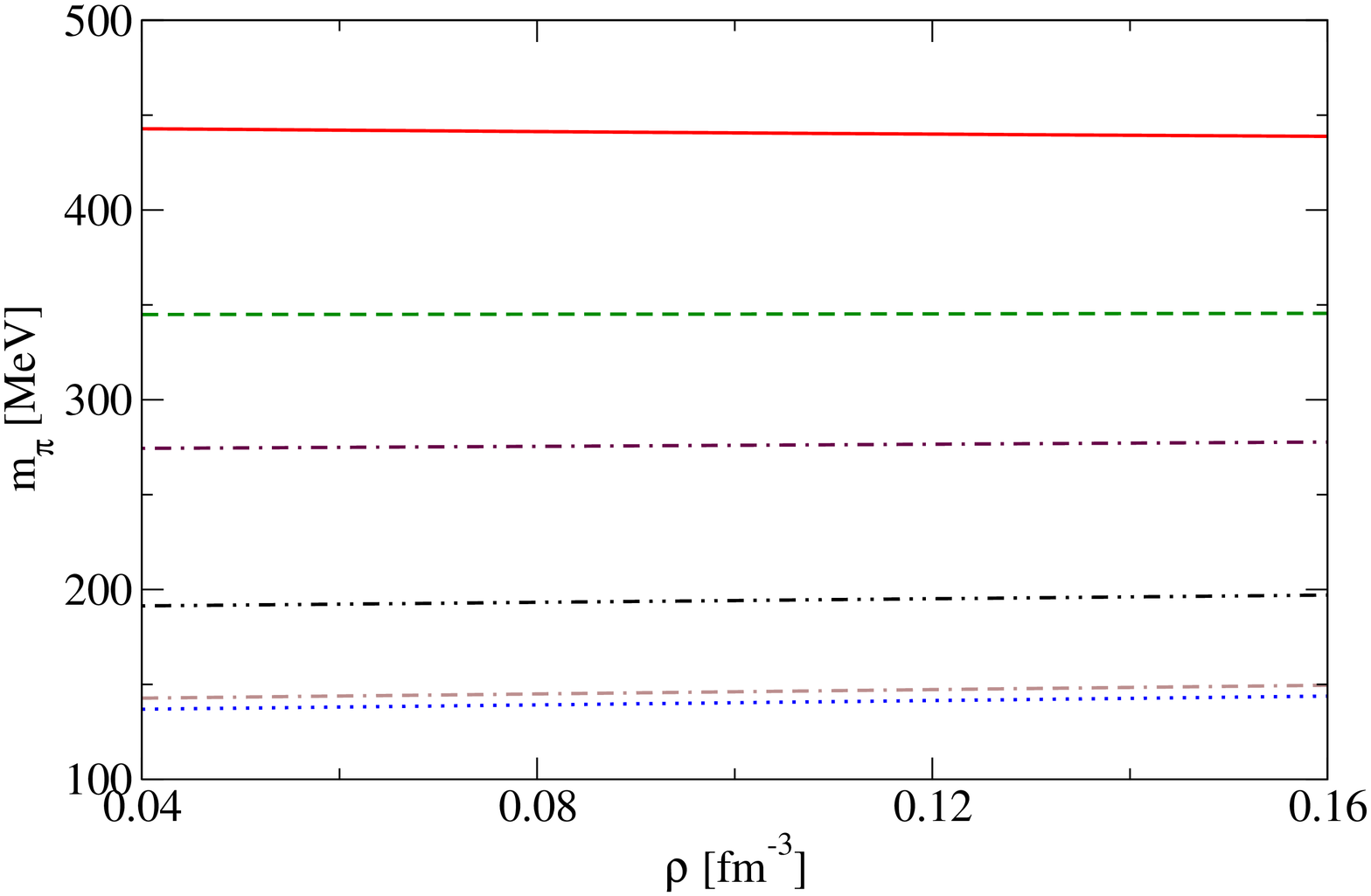}
 % feynman.eps: 0x0 pixel, 300dpi, 0.00x0.00 cm, bb=148 615 349 677
 \caption{(Color online) Upper panel: $\pi^-$ effective mass in a constant magnetic background, as a function of the density, 
 for zero temperature. From below, lines correspond to increasing values of the magnetic field, 
 namely $eB/ m_\pi^2= 0.01 , 0.1, 1, 3.2, 5.6, 10$.
 Lower panel: same as above, for $T = 50$ MeV.}
 \label{fig:eff_mass_plot}
\end{figure}

Fig.~\ref{fig:eff_mass_plot} shows the solution of Eq.~(\ref{eq:eff_mass_magnetic}) 
in the LLL approximation for the negatively charged pion, in the case of isospin symmetric 
nuclear matter, as a function of the density. The upper panel corresponds 
to the case of zero temperature. Clearly, the main contribution to the effective mass is given by 
the term proportional to $|eB|$ in Eq.~(\ref{eq:eff_mass_magnetic}) that represents the trivial Landau 
quantization shift. The variation with the density is almost negligible. 
Nevertheless, one can notice that in the case of low magnetic fields the effective mass slightly 
increases with the density, while in the case of extremely strong magnetic fields the mass decreases. 
Therefore, if we would plot only the self-energy contribution to the effective mass, 
it would exhibit an appreciable drop for high magnetic fields. We should, of course, stress that the extremal 
values of the magnetic field magnitude shown in Fig.~\ref{fig:eff_mass_plot} might bring some inconsistency
due to the fact that for too low fields one can not apply the LLL approximation, whereas for extremely strong 
fields one has to treat more carefully the chiral power counting. Yet, we believe that Fig.~\ref{fig:eff_mass_plot} 
illustrates clearly the qualitative trend as one comes from low to high magnetic fields.

\begin{figure}[htb]
 \centering
 \includegraphics[keepaspectratio=true,width=0.54\textwidth]{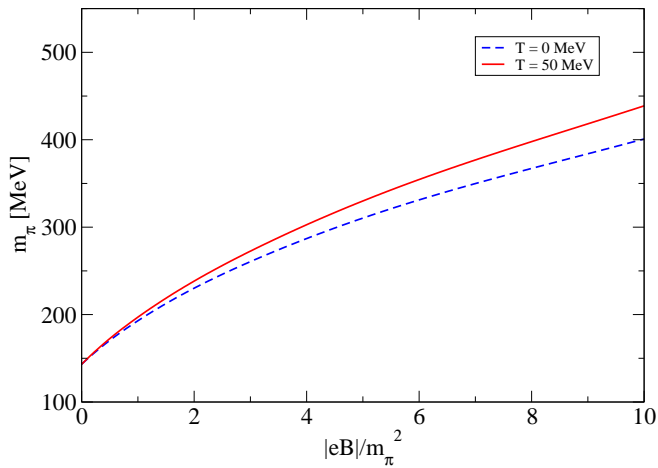}
 % feynman.eps: 0x0 pixel, 300dpi, 0.00x0.00 cm, bb=148 615 349 677
 \caption{(Color online) $\pi^-$ effective mass as a function of the magnetic field 
 at saturation density, $\rho = 0.16$ fm$^{-3}$. The (blue) dashed line is the result for
 the zero temperature case, while the full (red) line corresponds to a temperature $T = 50$ MeV.}
 \label{fig:eff_mass_vs_eB}
\end{figure}

In the lower panel the effect of the temperature is included. The main features of the plot remain the 
same but, at equivalent values of the magnetic field, one can see that these curves lie above the corresponding curves 
at zero temperature. This is also shown in Fig.~\ref{fig:eff_mass_vs_eB}, in which we fix the 
density at nuclear saturation, $\rho = 0.16$ fm$^{-3}$, and vary the magnetic field. 
The dashed curves correspond to the zero temperature case, while the full lines to $T=50$ MeV, and one can see 
that, for fixed density and temperature, the pion mass increases steeply with the external magnetic field. 

As remarked previously, the result for very strong magnetic fields ($|eB| \geq m_\pi^2$) 
has to be considered as an extrapolation. Indeed, for such high values of the magnetic 
field, the chiral power counting does not hold anymore. Since the low energy scale of the theory 
(the pion mass) and the hard scale become comparable, one should take into account higher order 
diagrams that become, in principle, relevant in this case. Nonetheless, the trend seems clear from 
within the region of validity of our approach and connects smoothly to the regions of higher and lower 
fields, which is encouraging.

\begin{figure}[htb]
 \centering
 \includegraphics[keepaspectratio=true,width=0.54\textwidth]{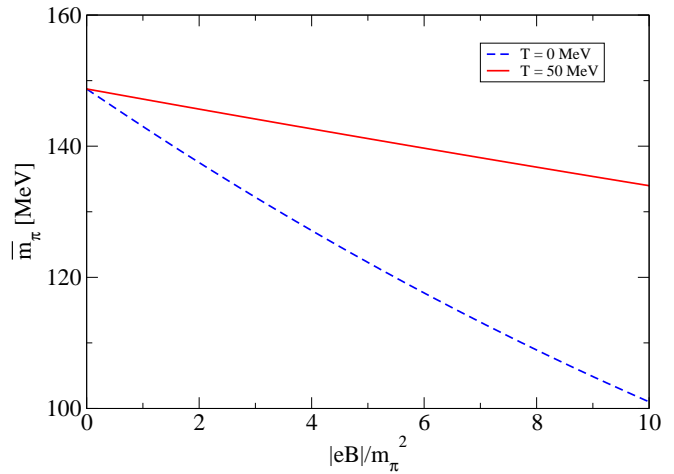}
 % feynman.eps: 0x0 pixel, 300dpi, 0.00x0.00 cm, bb=148 615 349 677
 \caption{(Color online) Solution of Eq.~(\ref{eq:eff_mass_only_self}) as a function of the magnetic field, 
 at saturation density $\rho = 0.16$ fm$^{-3}$. The (blue) dashed line is the result for
 the zero temperature case, while the full (red) line refers to a temperature $T = 50$ MeV.}
 \label{fig:self-contribution-m-vs-eB}
\end{figure}

To unveil the role played by the real part of the self-energy contribution to the mass of the $\pi^{-}$, we 
compute its effective mass having subtracted the trivial shift due to the presence of the
magnetic background (Landau quantization), namely we solve
\be\label{eq:eff_mass_only_self}
  \overline{m}_\pi^2 = m_\pi^2 - {\rm Re}~\Pi(\overline{m}_\pi^2,\vecq=0;\mathbf B)\,.
\ee 
Fig.~\ref{fig:self-contribution-m-vs-eB} displays our results for $\overline{m}_\pi$ as a function of the 
magnetic field, and shows a significant decrease in the effective mass of the $\pi^{-}$ as a function 
of the magnetic field. This effect is, of course, diminished as the temperature is increased.

The phenomenological motivation comes from physical systems with different charged particle species in 
the presence of moderately strong magnetic fields, as can be found {\it e.g.} in compact stars. In this case, 
one has to take into account the contributions coming from higher Landau levels, at least the first nontrivial 
ones. Thus, the trivial shift in the spectrum of different (charged) fermions and bosons will be of the same 
order ($\sim|eB|$), and the contribution that will be relevant for the behavior of the effective mass will possibly 
be determined essentially by the real part of the self-energy. In this context, we find that for the negatively charged pion 
the effective mass is lowered by the presence of a strong magnetic background. Due to this feature, properties 
of a system involving dense nuclear matter and leptons, as can be found in compact stars and supernovae~\cite{2008JPhG...35h5201I,2013PhRvD..87d3006P}, 
might change significantly, depending on the behavior of the spectrum of charged fermions. 

%%%%%%%%%%%%%%%%%%%%%%%%%%%%%%%%%%%%%%%%%%%%%%%%%%%%
\section{Summary}
\label{section:conclusions}

We have investigated hadronic matter in the low-energy sector in the presence of a strong magnetic field 
in the controlled setup of chiral perturbation theory. More specifically we have studied the modification of 
the pion self-energy at finite temperature due to its interaction with a low-density, isospin-symmetric nuclear 
medium in the presence of an external magnetic field. 

To one loop, for fixed temperature and density, we found that the pion effective mass increases steeply with 
the magnetic field, a result that is enhanced when temperature is included, as expected. As a function of the 
density, on the other hand, the behavior of the effective mass is quite flat for different values of the field. 
However, even keeping in mind the caveat of our method when considering too low or too high fields, it seems 
clear that there is a qualitative change in the overall behavior: augmentation of the effective mass for low fields 
and depletion for high fields. The latter effect is, of course, hampered as we increase the temperature.

A subtle point that can play a relevant role in some actual physical systems, with different charged particle species 
in the presence of moderately strong magnetic fields, is the fact that the increase in the effective mass of the $\pi^{-}$ 
with the magnetic field is due solely to the trivial Landau quantization shift $\sim |eB|$, since the real part of the 
self-energy is negative in this case. As was shown in the previous section, if we subtract the former trivial effect, 
the effective mass of the negatively charged pion drops considerably with the magnetic field. 
If all charged particles undergo an approximately equivalent trivial shift of this 
sort, the modifications that may be relevant, phenomenologically, are those brought about by the real part of the 
respective self-energies. In this case, we find that \textit{the pion mass decreases by $\sim10\%$ for a
magnetic field $|eB|\sim m_\pi^2$, which favors pion condensation at high density and low 
temperatures.} Such scenario may take place in neutron star matter and supernovae and requires further 
investigation.

%%%%%%%%%%%%%%%%%%%%%%%%%%%%%%%%%%%%%%%%%%%%%%%%%%%%
\section*{Acknowledgments}
The work of GC was supported by the HGS-HIRe graduate program 
at Frankfurt University. The work of ESF was financially supported by 
the Helmholtz International Center for FAIR within the framework of the 
LOEWE program (Landesoffensive zur Entwicklung Wissenschaftlich-\"Okonomischer 
Exzellenz) launched by the State of Hesse. GC thanks S. Beni\'c for useful discussions.

%%%%%%%%%%%%%%%%%%%%%%%%%%%%%%%%%%%%%%%%%%%%%%%%%%%%
% \bibliography{bib_chipt}{}
% \bibliographystyle{unsrt}

\end{document}